\documentclass[11pt]{article}
\usepackage{graphicx}
\usepackage{float}
\usepackage{wrapfig}
\usepackage{authblk}
\begin{document}
\title{On the elliptic locus of a family of projectiles}
\author{Joseph A Rizcallah\\joeriz68@gmail.com}
\affil{School of Education, Lebanese University, Beirut, Lebanon}
\date{}
\maketitle
\vspace{10pt}
\begin{abstract}
A little known property of free-fall motion is the elliptic locus of the maximum heights attained by coplanar projectiles launched from a single point in different directions with the same initial speed. Another, less known and perhaps somewhat surprising, property of this family consists in the said ellipse being also the geometric locus of the critical points of the projectiles' distances from the launch point. In the article, to gain a better perspective on the geometry involved, we consider these loci from the standpoint of a free-falling frame. It is shown that, in this reference frame, the considered loci are congruent circles tangent to each other along the horizontal and internally tangent to another, twice as big a circle, which represents the locus of the projectiles' ranges. This simplified geometric description is employed to recover the elliptic locus in the laboratory frame and further explore its properties.
\end{abstract}

\section{Introduction}
Projectile motion is an inalienable part of any introductory mechanics course. It also is a virtually inexhaustible source of inspiration for physics teachers and educators. In particular, families of projectiles exhibit some peculiar geometric properties, which have captured the attention of many a practitioner in the field. In this respect, of special interest is the family of coplanar projectiles launched from the same point in different directions with a constant initial speed. A well-known property of this family is the so called synchronous circular locus~\cite{dragellip}, which consists in simultaneously launched projectiles of the family forming a circle at any one instant of time. Another genuinely geometric, independent of time, feature of this family is the parabolic envelope that divides the vertical plane into an accessible zone and a safe domain~\cite{jean,me}.

Another less known geometric property of the aforementioned family is that the vertices of its member parabolas lie on an ellipse, which also turns out to be the locus of the critical points of the projectiles' distances from the launch point. In other words, every point of the said ellipse is, on the one hand, the apex of a parabolic trajectory of the family and, on the other, it is some other member's critical point at which the distance from the launch point admits a local maximum or minimum. In the contemporary literature, both these results were brought to light in~\cite{ellip}, though, it seems, they had been known earlier~\cite{noteellip}. Moreover, an undergraduate experiment~\cite{expellip} was devised to observe this ellipse, and the effect that linear air drag has on it was discussed~\cite{dragellip}. Recently, an alternate view on this elliptic property has been suggested in~\cite{altellip} (see also~\cite{mihas}).
   
In the paper, we offer yet another view on the elliptic locus associated with the above mentioned family and, in so doing, we attempt to illustrate the role of the reference frame in determining the shape of a geometric locus in general. We consider the motion of the projectiles under the assumptions of a uniform gravitational field $\vec{g}$ and negligible air resistance. Our treatment uses very little calculus and should be accessible to all physics students with some background in elementary plane geometry. The article is organized as follows: in the next two sections we introduce the problem and consider it from the standpoint of a free-falling reference frame, where the geometry proves to be particularly simple. In section~\ref{lab} we recover the elliptic locus by transforming the obtained loci in the free-falling frame back to the laboratory frame. Section~\ref{prop} explores some further properties of the elliptic locus using the simplified geometry of the corresponding loci in the free-falling frame. In the last section we wrap up our results and conclude the paper with a brief discussion of the adopted approach.

\section{Setting up the problem}
\label{set}
Consider a projectile shot from a fixed point (taken as the origin O of the coordinate system) with an initial speed $v$ at an angle $\theta$ to the horizontal. Keeping $v$ constant and varying $\theta$ in the range between $0$ and $\pi$ we obtain a family of projectiles launched with the same speed in the upper half of a fixed vertical plane. Throughout the paper we use $v/g$ and $v^2/g$ as units of time and length respectively. In these units the projectiles have a unit initial speed $v=1$ and a unit acceleration $a=1$, so their equations of motion are given by
\begin{equation}
\label{eqmotion}
x= t\cos \theta ,~~y = -\frac{1}{2}t^2+t\sin \theta,
\end{equation}
where, as usual, the directions $\hat{x}$ and $\hat{y}=-\hat{g}$ correspond to $\theta=0$ and $\theta=\pi/2$ respectively.

\begin{figure}
\centering
\includegraphics[scale=0.4, trim = 0 100 0 100, clip=true]{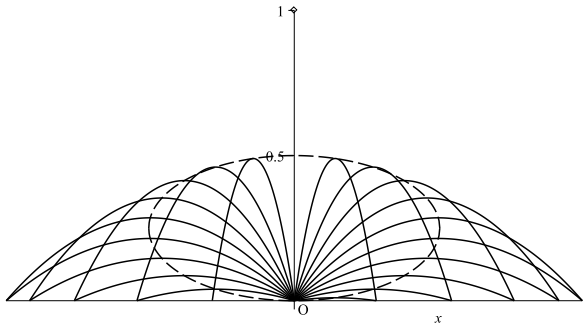}
\caption{A few members of the family of projectiles described in the text together with its elliptic locus (dashed) of maximum heights.}
\label{ell}
\end{figure}

Let us first consider the locus of maximum heights in the laboratory frame. Each projectile attains its maximum height at 
\begin{equation}
\label{tmaxh}
t=\sin\theta.
\end{equation}
Substituting for $t$ from~(\ref{tmaxh}) in the equations of motion~(\ref{eqmotion}) and subsequently eliminating the angle $\theta$, gives the elliptic locus
\begin{equation}
\label{ellip}
4x^2+16\left(y-\frac{1}{4}\right)^2=1.
\end{equation}
Figure~\ref{ell} shows ellipse~(\ref{ellip}) together with a few trajectories. Note that~(\ref{ellip}) is equivalent to equation (3) in~\cite{ellip}. The latter can be recovered from~(\ref{ellip}) if we revert to the conventional SI units by replacing $x$ and $y$ by $gx/v^2$ and $gy/v^2$ respectively. 

We see that deriving the elliptic locus of maximum heights in the laboratory frame is fairly straight forward. Turning to the locus of critical points of the distance from the launch point, one may be tempted to tackle the problem in the laboratory frame too. However, as we shall see below, a more elegant and illuminating solution can be found.

To begin with, note that the critical points are the local minimum and maximum of the projectile's distance from the origin. At these points the position and velocity vectors are perpendicular. In coordinate form, this condition reads $x\dot{x}+y\dot{y} = 0$, where an overdot denotes differentiation w.r.t. time $t$. Upon substitution from~(\ref{eqmotion}) and their time derivatives, after some manipulation, we get
\begin{equation}
\label{tcrit}
t^2-3t\sin\theta +2=0,
\end{equation}
where the trivial root $t=0$ has been ignored. One, in principle, can proceed by solving this equation for $t$ and substituting in~(\ref{eqmotion}), as before, try to eliminate the angle and obtain the equation of the locus, which happens to be the same ellipse~(\ref{ellip}). However, this method is not as straight forward as the one described above for the maximum heights and, more importantly, it does not explain the unexpected coincidence of the two loci. To gain a better understanding of the situation, we choose to transform the problem to a free-falling frame.

\section{Switching to a free-falling frame}
\label{fff}     
\begin{figure}[h]
\centering
\includegraphics[scale=0.4]{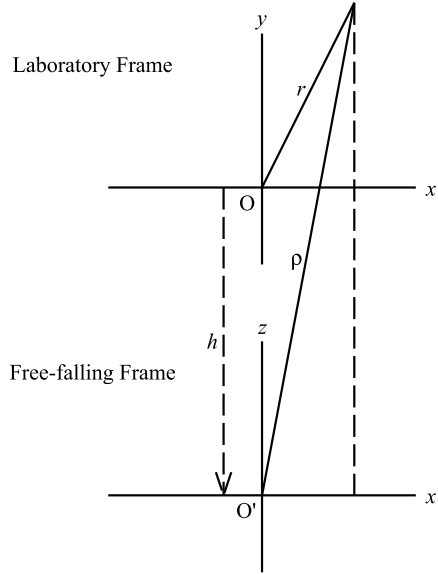}
\caption{The coordinate systems used in the laboratory and free-falling reference frames.}
\label{frames}
\end{figure}

Let us put ourselves in the shoes of a free-falling observer. More precisely, we assume that the free-falling observer starts his motion from rest simultaneously (at $t=0$) with the projectiles. As reckoned by this non-inertial observer, a projectile experiences no acceleration due to gravity and its motion is uniform with a unit velocity $\vec{v}$ in the direction $\theta$. Moreover, relative to this observer the ground together with the launch point perform uniformly accelerated motion in the 'upward' direction at a unit rate. If we assume that the free-falling observer and the laboratory observer use coordinate systems which coincide at $t=0$, then the observers will agree on the $x$-coordinate of a projectile, but generally disagree on its $y$-coordinate. Accordingly, we shall use $z$ and $\rho = \sqrt{z^2+x^2}$ to denote respectively the projectile's vertical coordinate and distance from the origin O$^\prime$ in the free-falling frame. Figure~~\ref{frames} shows the laboratory and free-falling reference frames at an arbitrary instant of time $t$.

\begin{figure}[h]
\centering
\includegraphics[scale=0.4]{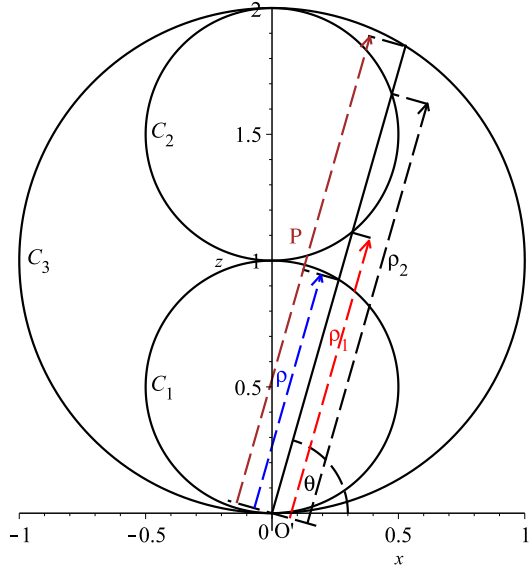}
\caption{The three circular loci associated with the considered family of projectiles in the free-falling frame. See text for details.}
\label{cir}
\end{figure}

Let us now consider our two loci from the standpoint of the free-falling observer. We begin with the locus of maximum heights. In the free-falling frame and in the adopted units, a projectile preforms uniform motion at a unit speed, so its distance $\rho$ from the origin O$^\prime$ is given by 
\begin{equation}
\label{fffdist}
\rho=t.
\end{equation}
At this point it is worth noting that while in the laboratory frame the point of maximum height has a geometric and a physical significance, as the point of maximum curvature and least speed, in the free-falling frame it is just another point along the straight trajectory at which the ground is farthest 'below' the projectile. Notwithstanding, the time the maximum height is attained is sill given by~(\ref{tmaxh}). Together with~(\ref{fffdist}) this gives
\begin{equation}
\label{circmaxh}
\rho=\sin\theta,
\end{equation}
for the locus of maximum heights in the free-falling frame in the adopted units. This is readily recognized to be the polar equation of a circle $C_1$ of radius $1/2$ and center $(x=0,z=1/2)$, as shown in figure~\ref{cir}. In the Cartesian coordinates $x,z$ this circle has the equation 
\begin{equation}
\label{circ1}
x^2+\left(z-\frac{1}{2}\right)^2=\frac{1}{4}.
\end{equation}
      
We now turn to the locus of the critical points of the projectile's distance from the launch point. According to~(\ref{tcrit}) the times $t_1$ and $t_2$, when defined, at which the projectile attains its maximum and minimum distance from the launch point have a constant product $t_1t_2=2$, independent of the launch angle $\theta$. Together with~(\ref{fffdist}) this implies
\begin{equation}
\label{critcirc}
\rho_1\rho_2=2.
\end{equation}
It is a well-known fact of geometry~\cite{aeg} that~(\ref{critcirc}) defines a circle $C_2$ (see figure~\ref{cir}) w.r.t. which the origin O$^\prime$ has a power of $2$. It is easy to see that this circle has a radius equal to the maximum of $|\rho_{2}-\rho_{1}|/2$. To find this radius, note that together with the identity $\left(\rho_2 - \rho_1 \right)^2 = \left(\rho_2 + \rho_1 \right)^2 - 4\rho_1\rho_2$ equation~(\ref{critcirc}) implies that the difference $|\rho_2-\rho_1|$ and the sum $\rho_2+\rho_1$ attain their maxima simultaneously. On the other hand, from equations~(\ref{tcrit}) and (\ref{fffdist}) we have $\rho_2+\rho_1=3\sin \theta$ with a maximum of 3 corresponding to $\theta = \pi/2$. It follows that the sought for radius is $1/2$ and that the center of $C_2$ is at $(x=0,z=3/2)$, so that its Cartesian equation reads 
\begin{equation}
\label{circ2}
x^2+\left(z-\frac{3}{2}\right)^2=\frac{1}{4}.
\end{equation}
One readily sees that the circle $C_2$ touches the circle of maximum heights $C_1$ along the horizontal at the point $(x=0,z=1)$, as depicted in figure~\ref{cir}.

\begin{figure}[h]
\centering
\includegraphics[scale=0.4]{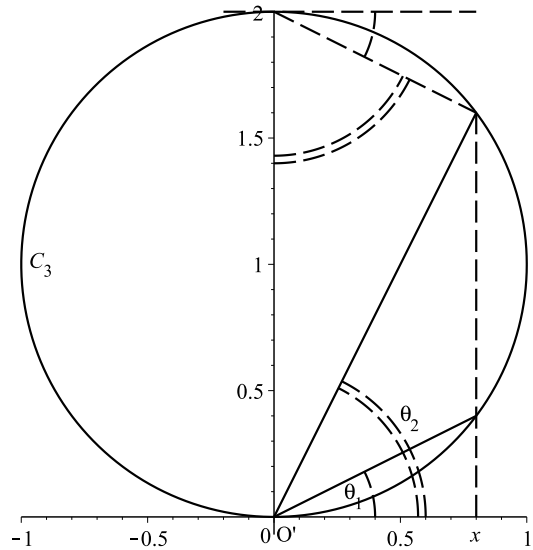}
\caption{The congruence of the indicated angles implies $\theta_1+\theta_2=\pi/2$ for two projectiles with the same range $x=R$.}
\label{cir3}
\end{figure}

It will be expedient to consider one more circle $C_3$, the image of $C_1$ under a dilation of center O$^\prime$ and ratio 2. This circle is the geometric locus of the projectiles' ranges in the free-falling frame. To see this, suffice it to recall that a projectile's flight time is twice the time it takes to reach its maximum height, i.e. $t=2 \sin \theta$ for the range. So by~(\ref{fffdist}), the distance $P$ from O$^\prime$ to a point on $C_3$ is given by $P= 2 \sin \theta$, which upon comparison with~(\ref{circmaxh}) yields the desired result $P=2\rho$. Figure~\ref{cir} depicts all the three circles described above. In passing, we note that the well-known complementarity of the two launch angles corresponding to a range $x=R$ follows directly from symmetry and the properties of inscribed angles in $C_3$, as illustrated in figure~\ref{cir3}.

\section{Switching back to the laboratory frame}
\label{lab}
The goal of this section is to provide an alternative proof of the elliptic locus by showing that $C_1$ and $C_2$ transform into ellipse~(\ref{ellip}) in the laboratory frame. To this end, we note that for any projectile the $y$-coordinate and $z$-coordinate at an artitrary time $t$ are related by
\begin{equation}
\label{transf}
y=z - h,
\end{equation}
where $h=1/2gt^2$ is the distance the free-falling frame descends in time $t$, as shown in figure~\ref{frames}. 

Consider first the transform of $C_3$ into the laboratory frame. Since $C_3$ is the locus of the projectiles' ranges in the free-falling frame, its transform to the laboratory frame must be a segment of the $x$-axis with ends at the maximum range on either side of the origin O, i.e. $y=0$ and $-1 \leq x \leq 1$. By~(\ref{transf}), this implies  
\begin{equation}
\label{transc3}
Z = H,
\end{equation}
where $Z$ and $H$ are respectively the $z$-coordinate and the descent distance of the free-falling frame corresponding to an arbitrary point on $C_3$.

\begin{figure}[ht]
\centering
\includegraphics[scale=0.4]{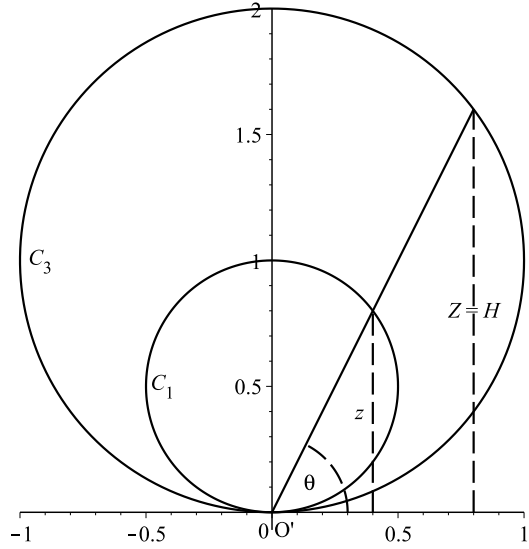}
\caption{To find the transform of $C_1$ we make use of $z=Z/2$ and $Z=H$ for a point on $C_3$. See text for details.}
\label{cir1}
\end{figure}

We are now ready to consider the transforms of $C_1$ and $C_2$. We shall start with the former. For any point on $C_1$, by the definition of $C_3$ we have $\rho=P/2$. Since $z$ scales as $\rho$ it follows that $z=Z/2$ (see figure~\ref{cir1}). Moreover, by~(\ref{fffdist}) $z$ also scales as $t$, while the descent distance $h$ scales as $t^2$, whence $h=H/4$ for any point on $C_1$. Combining these results for $z$ and $h$ with~(\ref{transf}) and~(\ref{transc3}), we obtain $y=z/2$ for any point on $C_1$. So the transformation of $C_1$ from the free-falling frame to the laboratory frame involves a compression of ratio $1/2$ and the $x$-axis as its fixed axis. It follows that, in the laboratory frame, the circle $C_1$ is squashed into the ellipse~(\ref{ellip}). To algebraically confirm this result, one replaces $z$ by $2y$ in (\ref{circ1}) and multiply through by $4$ to obtain~(\ref{ellip}). It is worth noting that this compression can be traced back to the well-known fact (see e.g. \cite{maxheight}) that the projectile's maximum height is half the altitude it would attain during the same time if gravity were turned off. Also note that the circle $C_1$ is identical to that used in reference~\cite{altellip}. In our approach, however, this circle arises naturally without the need of extraneous elements, such as mirrors, and acquires a physical and geometric significance of its own, namely as the locus of maximum separation between the projectiles and the ground in the free-falling frame.

\begin{figure}[ht]
\centering
\includegraphics[scale=0.4]{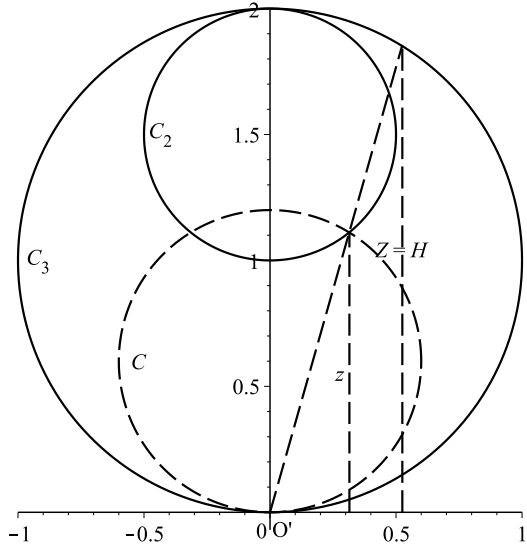}
\caption{The auxiliary construction used to determine the image of $C_2$ in the laboratory frame. See text for details.}
\label{cir2}
\end{figure}

We now turn to the transform of $C_2$. To this end, consider the image $C$ of $C_3$ under a dilation of center O$^\prime$ and ratio $k$, where $1/2 \leq k \leq 1$ (see figure~\ref{cir2}). By the aforementioned scaling of $z$ and $h$, a point on $C$ has $z = kZ$ and $h=k^2H$, so by~(\ref{transf}) and (\ref{transc3}) its $y$-coordinate in the laboratory frame is given by $y=(1-k)z$. Now, the circle $C$ intersects $C_2$ at two points with equal $z$. Using the Cartesian equation of $C$, i.e. $x^2+(z-k)^2=k^2$, along with~(\ref{circ2}), with some algebra one finds $z = \frac{2}{3-2k}$ for the intersection points. Eliminating $k$ between the last two results, yields $y = -z/2 +1$. This can be interpreted as a composition of the following three transformations performed on $C_2$: a reflection about the $x$-axis together with a compression of ratio $1/2$ w.r.t. the same axis followed by a unit translation in the positive vertical direction. It is readily seen that the image of $C_2$ under this transformation is again ellipse~(\ref{ellip}). Algebraically, this means that ellipse~(\ref{ellip}) is obtained from~(\ref{circ2}) too by substituting for $z$ from $y = -z/2 +1$ (and multiplying through by $4$). Note, however, that due to reflection, the upper and lower semicircles of $C_2$ are mapped to the ellipse's lower and upper halves respectively.

\section{Further properties of the elliptic locus}
\label{prop}
Now time has come to reap the benefits of our work. In this section we discuss three properties of the family under consideration, which are made almost obvious by the effected transformation to the free-falling frame.

The first property consists in the fact that none of the projectiles' critical points occurs before {\it all} the projectiles in the family have attained their maximum heights. It is easy to see that for a single projectile the critical points (if any) occur after the maximum height is reached, but it is not immediately clear that this must hold for the entire family. This property follows from figure~\ref{cir}, which makes it evident that any point on $C_2$ is farther away from the origin (hence occurs later) than any point on $C_1$. This can be made observable if we picture every projectile in the family is equipped with a clock mechanism that is set to emit a red flash when the maximum height is reached and a green flash when a critical point is passed. Then an observer in the laboratory frame would see ellipse~(\ref{ellip}) traced out twice: first in red light from bottom to top then in green light from top to bottom. Note that the clock must be set in advance, as the projectiles experience no change in acceleration that could trigger the flashes.

\begin{figure}[ht]
\centering
\includegraphics[scale=0.4]{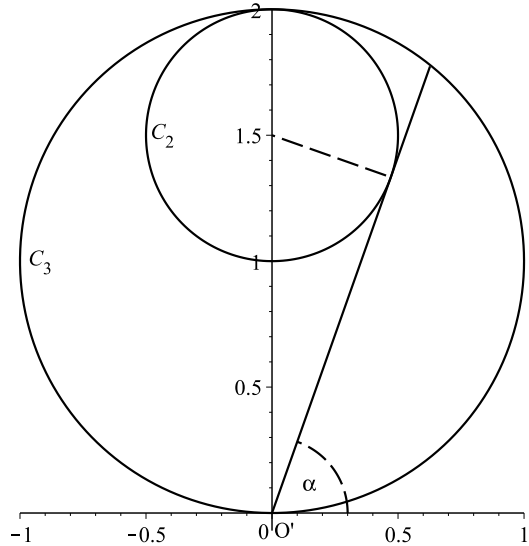}
\caption{The line from O$^\prime$ tangent to the circle $C_2$ corresponds to the singular trajectory that touches ellipse~(\ref{ellip}) in the laboratory frame. The launch angle this line makes with the $x$-axis is the special angle $\alpha$ we are looking for.}
\label{cirtan}
\end{figure}

The second property has to do with two special angles associated with the 'singular' trajectory that touches ellipse~(\ref{ellip}). In general, a projectile, on its way down, either cuts the ellipse~(\ref{ellip}) twice or not at all; one only projectile has a trajectory touching that ellipse at one (doubly critical) point. In the free-falling frame, the latter corresponds to the tangent line drawn from the origin O$^\prime$ to $C_2$, with $\rho_1=\rho_2=\sqrt{2}$ by~(\ref{critcirc}). The launch angle $\alpha$ corresponding to this projectile as well as the angle $\beta$ at which the point of tangency hangs above the horizontal in the laboratory frame are the special angles we have in mind. From figure~\ref{cirtan}, we see that $\cos\alpha= \sin(\pi/2-\alpha) =1/3$, so $\alpha = \arccos(1/3) \approx 70.53^\circ$. To find $\beta$, note that for the point of tangency we have $x=\sqrt{2}\cos\alpha=\sqrt{2}/3$ and $z=\sqrt{2}\sin\alpha=4/3$ in the free-falling frame, so by $y=-z/2+1$ of the previous section, in the laboratory frame, we have $y=1/3$ and $\beta = \arctan(y/x) = 1/\sqrt{2}=35.26^\circ$. 

Physically, projectiles launched at angle $\alpha$ or shallower will never approach the launch point, whereas those launched at a steeper angle approach the launch point on their way down while inside the ellipse~(\ref{ellip}). Note that the time spent inside the ellipse is given by $|\rho_2-\rho_1|=\sqrt{9\sin^2\theta-8}$, which vanishes for $\theta=\alpha$. Also note that neither $\alpha$ nor $\beta$ (nor any other angle associated with projectiles, see~\cite{ang1,ang2}) depend on the initial speed or acceleration due to gravity, the reason being that no dimensionless quantity can be formed out of these parameters.

\begin{figure}[h]
\centering
\includegraphics[scale=0.4]{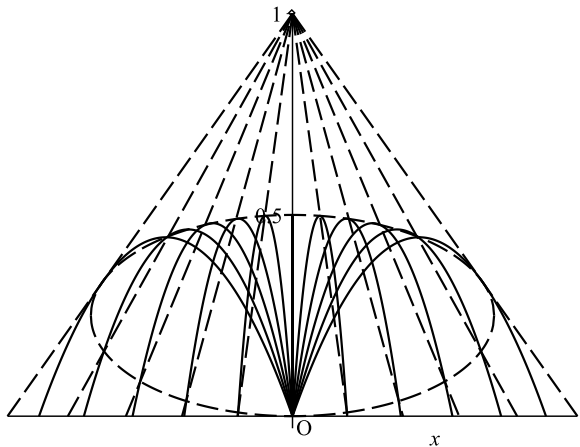}
\caption{Every dashed line is drawn through a trajectory's pair of critical points located on ellipse~(\ref{ellip}). All these straight lines converge at a single point on the $y$-axis.}
\label{prop2}
\end{figure}

The last feature is a mere geometric curiosity. From figure~\ref{cir} we see that for a given projectile the critical points, if any, are collinear with the origin O$^\prime$. This collinearity is preserved by the linear transformation $y=-z/2+1$ from the free-falling frame to the laboratory frame. Therefore, for any parabolic trajectory in the family, the critical points, if any, are collinear with the image of O$^\prime$ in the laboratory frame, i.e. with the point at $(x=0,y=1)$. This is illustrated in figure~\ref{prop2}. This property makes it relatively easy to determine the above mentioned (doubly critical) point of tangency, as that where a straight line through $(0,1)$ touches ellipse~(\ref{ellip}).

\section{Conclusion}
In the article, we consider a family of projectiles launched from the same point with a constant speed. We derive in a new way the elliptic locus for the maximum heights and the critical points of their parabolic trajectories, point out the time ordering of those, and note a geometric curiosity about the critical points. We also add to the list of launch angle special values a previously overlooked one associated with the 'singular' trajectory with a doubly critical point. 

Our method involves a simplification of the geometry, achieved as a result of transforming the problem into a non-inertial free-falling frame. The benefits of this method, are the simplicity of geometry as well as the ease of the mathematical analysis. Its downside, perhaps, is the not so intuitive description of the relative motion required in the free-falling frame.

Another advantage of this approach is that it explicitly demonstrates that the coincidence of the loci of maximum heights and critical points is a frame-dependent effect rather than a genuine physical one. More generally, our treatment makes it clear that the geometric shape of a non-synchronous locus, i.e. a geometric place of positions at different times, may change from one frame to another and is thus devoid of a clear physical meaning.

\end{document}